\documentclass[twocolumn]{revtex4}

\usepackage{graphicx}

\begin{document}

\title{Cosmological dinosaurs}

\author{V.~K. Dubrovich}
\affiliation{St. Petersburg Branch of Special Astrophysical Observatory,
Russian Academy of Sciences, 196140, St. Petersburg, Russia}
\affiliation{Nizhny Novgorod State Technical University n. a. R. E. Alekseev,
LCN, GSP-41, N. Novgorod, Minin str. 24, 603950}
\author{S.~I. Glazyrin}
\email{glazyrin@itep.ru}
\affiliation{Institute for Theoretical and Experimental Physics, Moscow, Russia}

\begin{abstract}
  The hypothesis of existence of primordial black
  holes with large masses ($\geq 10^6 M_\odot$), formed at the earliest
  stages of the Universe evolution, is considered in the paper. The
  possibility does not contradict some theories, see e.g. \cite{BarkanaLoeb_PhysRep_2001}, and may match
  new observational data. In particular, this scenario of evolution
  could describe some peculiarities in distant galaxies and
  quasars. Calculations of evolution of central body mass in
  protogalaxies for different initial conditions are presented. It is
  shown that the sufficient rate of BH mass growth is not achieved in
  the standard scheme without complex additional
  assumptions. Moreover, the appearance of a primordial black hole in
  the epoch of primordial nucleosynthesis could significantly change
  the chemical composition around it. This can lead to different
  exotic stars with low mass and nonstandart metals enrichment. The
  proposed scheme is not considered as universal. On the other hand,
  if only tiny part of existed objects have the considered nature, it
  gives a unique possibility to study extremal stages of matter and
  fields evolution in our Universe.
\end{abstract}

\maketitle

\section{Introduction}

This paper considers the possibility of existence of rare
high-redshift objects: galaxies and quasars, formed much earlier than
the typical composition of the Universe. We also propose the mechanism
of the formation of such objects.

Today the theory of formation of the structure of the Universe
is well confirmed by observations. It describes the whole history of our
world evolution. Observations cover only tiny part of it: they are
limited now to the redshift range
$0<z<10$ (see refs. \cite{LehnertEtAl_Nature_2010,Bouwens_Nature_2011})
and the point $z\approx 1000$ (CMB observations).
Modern and future missions and experiments have the
possibility to expand the region towards the moment of the Universe
creation. On this way we are discovering now more and more distant galaxies
with no end in sight. The theory tells us that everything around was built in
evolutionary way: for some period after recombination there was no source of light in
the Universe -- Dark Ages, firsts stars appeared approximately at
$z\approx 30$, and only further galaxies were constructed. So the
observational trend of discovering new objects (galaxies and quasars) should be violated at
some redshift. At this point we expect the discovery of something
beyond standard predictions.

But what if we find a galaxy so old that it lies beyond the redshift
region allowed by the standard theory (ST)? Then something in our
understanding needs changing. We consider the theory to be
generally correct, that is the number of such extraordinary objects is
negligible. But it requires small additions to be applied.
We propose primordial big black holes (PBBHs) as simple candidates to such
extension. By this term we mean black holes with masses $M>M_\odot$
formed in the early Universe (before recombination). In this case we only change the initial conditions for
the problem of structure formation. The result could account for the early formation
 ($z_{\rm formation}\gg 30$) of rare objects (we propose to call them
``cosmological dinosaurs''). We state that it may be not so fantastic:
recently two galaxies with unusually high black
hole-to-bugle mass ratio were discovered
by \cite{BogdanFormanEtAl_1203.1641}, and a discovery of a star with very low
metallicity, see \cite{CaffauEtAl_Nature_2011}. These examples offer
difficulties for the theory. They are candidates for ``dinosaurs''.

The structure of the paper is the following. In the Section \ref{sec:st}
the standard theory of the structure formation is considered along
with some its bottlenecks. The Section \ref{sec:SMBH} considers the question of
supermassive black holes growth. We show that the creation of a SMBH
evolutionary (and after recombination) is not a simple problem even
for known today quasars. We consider their masses as a simple general
criterium for the dinosaurs: too massive BH in the early Universe
could not be described by the standard theory.
The general aim of this article is to introduce the concept of
cosmological dinosaurs and show that there could be need in
some cases. Further works will consider this possibility more carefully.

\section{The standard theory of structure formation}
\label{sec:st}

Let us consider the history of the Universe along lines of
the standard theory, for good review see \cite{BarkanaLoeb_PhysRep_2001}. The first relevant
stage is inflation. It is inflation that generated fluctuations of
matter density and gravitational waves responsible for the creation of
all astronomical objects. The earliest
experimental evidence of these fluctuations available today are
observations of CMB temperature spatial distortions. They reflect the state of the Universe at the
epoch of the recombination, for recent results see \cite{KomatsuDunkleyEtAl_ApJS_2009} (the future
relic neutrino telescopes will make it possible to observe
the Universe at the moment much earlier, $t\approx 1$ s after the Big Bang). According to
them the density variations at $z\approx 1000$ were
at the scale of $\delta\rho/\rho\sim 10^{-5}$. The standard theory
assumes existence of only these
perturbations, which are quite uniform.

After the recombination the general contribution to the evolution of
the Universe is created by matter,
which is composed of DM and baryonic matter.
The difference between two components is due to existence of
thermodynamic pressure in baryonic
component. It leads to the notion of the Jeans wavelength for baryons:
\begin{equation}
  \lambda_J=\left(\frac{\pi c_s^2}{\rho_b G}\right)^{1/2}.
\end{equation}
This length separates oscillatory and exponential growth of linear density
perturbations. In case of efficient cooling the Jeans mass $M_J\sim
\rho\lambda_J^3$ defines the mass of an object that contracts and
becomes gravitationally bound.

The reality is a slightly more complex. The cold dark matter collapses
first and creates potential wells. Baryonic matter accretes into
these potential wells. Cosmological hydrodynamic simulations like
in \cite{SpringelEtAl_Nature_2005} are required to fine reconstruct the
process of structure formation. The simple analytic consideration
from \cite{BarkanaLoeb_PhysRep_2001} shows that Jeans mass introduced earlier and a minimum halo mass in potential wells agree quiet well.


Together with the first collapsed objects low metallicity PopIII
stars appeared. These stars are quiet massive $M\sim 10^2\div 10^3
M_\odot$ with small lifetime, see
\cite{WoosleyHegerWeaver_RMP_2002,YoshidaBrommHernquist_ApJ_2004}.
Their appearance considerably change the evolution of the ambient
matter. In the absence of metals (and low temperatures $T<10^4$~K) the
molecular hydrogen is the most efficient coolant. First stars on the
one hand destroyed H$_2$ with their radiation, on the other hand they
started to enrich the Universe with metals ($Z>2$).
Powerful supernova explosions, a typical final stage of PopIII stars
evolution, spreaded these elements all over the Universe.

As was mentioned earlier only sophisticated hydrodynamical simulations
which take into account star formation and feedback effects
could give quantitative answers on galaxies
formation for redshifts when nonlinear evolution starts (see
\cite{SpringelEtAl_MNRAS_2005}). Recent results for the standard
theory could be found in works by several groups:
\cite{GreifJohnsonEtAl_MNRAS_2008,BrommYoshida_ARAA_2011,DiMatteo_ApJ_2012}.

Results of cosmological observations can be gathered in the
following statements.
The candidate for the most distant galaxy known today is located at $z\approx 10$
\cite{Bouwens_Nature_2011}, but the resolution do not permit yet to
determine exact parameters of this object. Most distant
confirmed galaxy is at $z=8.6$ \cite{LehnertEtAl_Nature_2010}.
And at $z=7\div 8$ there are a lot of galaxies with well known
characteristics presented in \cite{FinkelsteinEtAl_ApJ_2010}.
It can be seen from mentioned works that the theory satisfy these observational ``restrictions''.

We can't rule out the case that future observations may find objects
that are not confirmed by the standard theory. We can state that these objects,
if exist are very rare. To account for such probable discoveries there are
two variants of explanations: peculiar point(s) in initial perturbations
or our poor understanding of nonlinear evolution of these perturbations.

Let us start with the second variant. The chief process that
determines the speed of structure formation is cooling. In the regions
with intensive cooling objects are formed earlier. We believe that we
know all mechanisms of cooling and relevant physics, so for efficient cooling we
need high temperatures $T>10^4$~K or abundance of
metals. Immediately the question arises: how this region appears? But
from the point of view of the evolutionary hypothesis of our world appearance, we cannot
create such cosmological peculiarity without peculiarity in
initial conditions. So we came to the first variant.

According to the inflationary scenario the Universe was initially
created at the Planck length $l_{\rm P}$, which was then exponentially
expanded during the inflation stage (this is the general argument
explaining homogeneity). There is a variety of inflationary scenarios
each with own consequences for our world. Because of the absence of
observational experiment in that region of cosmic time everything
that satisfy only CMB restrictions could be created. Possibilities
for peculiar conditions are
wide: from nontrivial topological solutions to new types of particles.
These ``peculiar objects'' should satisty the following conditions 
not to be forgotten and influence the evolution of parts of the
Universe at $z\sim 10\div 100$.
This object should be massive enough $\ge M_\odot$ or have the
possibility to grow in mass. It should be long living $t\sim
t_{\rm recomb}$, otherwise the energy deposition by its decay will be spread over
large region of space by photons and became non-significant.

The most suitable candidates for initial conditions
perturbations are primordial big black holes. The possibilities of
their creation have been considered in many papers,
e.g. \cite{DolgovSilk_PRD_1993, RubinEtAl_SovJETP_2001, DokuchaevEtAl_AstronRep_2008, DolgovEtAl_NuclPhysB_2009},
for a review see \cite{Khlopov_ResAstronAstrophys_2010}. The key advantage of BH is
their possibility to grow in mass and size. It make possible for them
to influence large spatial volume. Also the
nucleosynthesis near BHs will proceed differently with bigger yield of
metals. It lead to faster cooling during the Dark Ages and earlier
matter collapse. From our point of view this is the best and the least
exotic candidate that could account for ``dinosaurs''.

\section{Supermassive black hole formation}
\label{sec:SMBH}

Let's consider the question of the supermassive black hole formation
move carefully. Their existence at definite redshift is a very simple
and robust criterium for the validity of the ST. The most distant quasar known today is located at
$z=7.085$ with a mass
$M\approx 2\times 10^9 M_\odot$, see \cite{MortlockWarrenEtAl_Nature_2011}. We will show that this fact creates
some difficulties for the theory that accounts for explanation of
the structure formation.

We will make some analytical estimations here and discuss the
growth of a hypothetical black hole.
The rate of matter falling on the BH could be written in general as
\begin{equation}
  \dot M_{\rm BH}=\mu m_p n \sigma v_{\rm matter},
  \label{eq:dMBHdt_general}
\end{equation}
where $\mu$ -- is an average atomic weight of matter, $v$ -- its
velocity, $n$ -- concentration of matter around the BH, $\sigma$ -- cross-section of capture.
In case matter is virialized, that is its velocity coincides with the
thermal energy and is described by the temperature $k_BT_{\rm vir}\sim
\mu m_p v^2$.
For this case the cross-section of capture is defined by the Bondi
radius:
\begin{equation}
  R_{\rm Bondi}=\frac{\mu m_p G M_{\rm BH}}{k_B T}.
\end{equation}
As a result the accretion on the black hole is described by the
Bondi--Hoyle formula from \cite{Bondi-1969}:
\begin{equation}
  \label{eq:bondihoyle}
  \dot{M}_{\rm Bondi}=\frac{\alpha 4\pi G^2 M_{\rm BH}^2m_H n}{c_s^3},
\end{equation}
where the dimensionless parameter $\alpha=(3\gamma^3)^{1/2}/4 = 0.93$.

The crucial parameter for the accretion rate is $n$.
From trivial consideration of a halo with mass $M_{\rm halo}$ and
radius $R_{\rm halo}$ it is equal:
\begin{equation}
  n=\frac{3}{4\pi R_{\rm halo}^3}\frac{M_{\rm halo}}{\mu m_H}=4.6\times
  10^{-2} \mu^{-1}R_{\rm halo,6}^{-3}M_{\rm halo,9}~{\rm cm}^{-3}.
  \label{eq:n_halo}
\end{equation}
The halo mass is in units of $10^9 M_\odot$, the radius is in units 6 kpc.
As we will see later this value is too low for the explanation of
efficient BH growth.
This estimation differs greatly from the real density in halo center in some cases.
Following \cite{MoMaoWhite_MNRAS_1998} a halo is considered composed
of DM spherical isothermal halo and a disc of baryon matter. The
latter is supported by the angular momentum of the system. In this
case the density of particles in the centre of halo increases by
several orders. The numerical value is presented in \cite{VolonteriRees_ApJ_2005}:
\begin{equation}
  n = 6\times 10^4 f^2_{d,0.5} \lambda^{-4}_{0.05} T_{{\rm
      gas},8000}^{-1}R_{{\rm vir},6}^{-4}M_{h,9}^2~{\rm cm}^{-3}.
  \label{eq:n_VolonteriRees}
\end{equation}
Here $f_d$ is the fraction of gas in the disc (normalized to 0.5) and
$\lambda$ is the spin parameter in units of 0.05.
The latter value should be considered accurately. In demonstrates that
a halo gave an angular momentum. The initial
cosmological perturbations are vortex free and the general question
arises: how fast do $\lambda$ grow (together with discs). The work \cite{OhHaiman_ApJ_2002} states fast
disc creation for the case of atomic cooling: when the temperature
exceeds $T>10^4$~K atomic cooling became so efficient that it keeps
$T=10^4$~K and satisfy $t_{\rm cool}<t_{\rm dyn}$, these conditions
lead to fast near isothermal gas collapse. The condition of the disc
stability $\lambda>\lambda_{\rm crit}$ diminishes low values of
$\lambda$. According to \cite{WarrenEtAl1992} the average value of the spin
parameter is $\overline\lambda=0.05$. We will use approximation
(\ref{eq:n_halo}) for $n$ when $T<10^4$~K and
(\ref{eq:n_VolonteriRees}) when $T>10^4$~K (setting then by hands the
temperature $T=10^4$~K). But we consider this fact to be carefully
analysed.

To make an evaluation of BH growth we should know the dependence of
$M_{\rm halo}(z)$. To calculate all parameters of
halos we will use the Press--Schechter formalism, see
\cite{BarkanaLoeb_PhysRep_2001}. First step is the calculation of
the variance
\begin{equation}
  \sigma^2(M)=\sigma^2(R)=\int\limits_0^\infty\frac{dk}{2\pi}k^2P(k)
  \left[\frac{3j_1(kR)}{kR}\right]^2,
  \label{eq:sigma_M_int}
\end{equation}
where $M=4\pi\rho_m R^3/3$ and $\rho_m$ is the mean density of matter
in the Universe,
$P(k)$ is a power spectrum of density perturbations. We will use the
approximation for this spectrum proposed by
\cite{EisensteinHu_ApJ_1999} (the paper is accompanied by the code to
calculate the integral (\ref{eq:sigma_M_int})).

\begin{figure}
  \centering
  \includegraphics[angle=270,width=\linewidth]{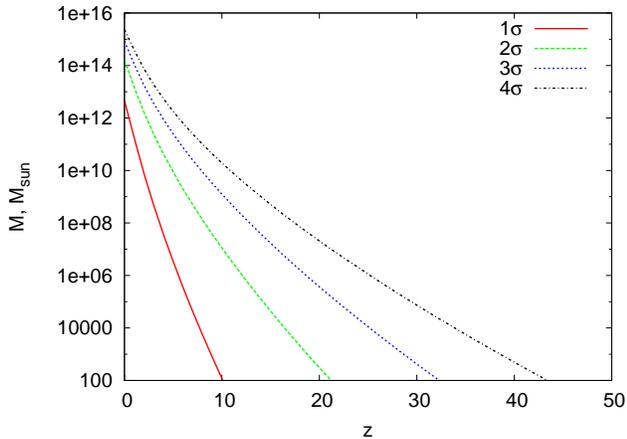}
  \caption{Halo mass vs redshift according to Press--Schechter model
    for different fluctuations.}
  \label{fig:M_z}
\end{figure}

Using the growth factor $D(z)$ (normalized as $D(0)=1$, so for clearance we omit
constant factors in $D_1$ definition):
\begin{equation}
  D(z)=\frac{D_1(z)}{D_1(0)},\quad D_1(z)=g(z)^{1/2}\int\limits^z
  \frac{1+z'}{g(z')^{3/2}}dz',
\end{equation}
\begin{equation}
  g(z)=\Omega_0(1+z)^3+(1-\Omega_0-\Omega_\Lambda)(1+z^2)+\Omega_\Lambda,
\end{equation}
the critical overdensity of a collapsed cloud can be calculated according to the top-hat model:
\begin{equation}
  \delta_{\rm crit}=\frac{1.686}{D(z)}.
\end{equation}
The solution of the equation
\begin{equation}
  \delta_{\rm crit}(z)=n\sigma(M)
\end{equation}
gives the dependence $M(z)$ for the collapsed cloud with the
$n$-$\sigma$ Gaussian fluctuation.
This model is very simplified but quite exactly reproduces history
of halo merges. The result of calculation is presented on Fig. \ref{fig:M_z}.
We will use this dependence as a basis for analytical estimation
of BH growth.

The Bondi--Hoyle accretion rate
(\ref{eq:bondihoyle}) is limited by some physical effects that
reduce the rate of accretion. First the Bondi radius should not exceed
geometrical parameters of halos:
the halo virial radius. The latter is calculated using known $M_{\rm halo}$:
\begin{equation}
  R_{\rm Bondi}<R_{\rm vir} = \left(\frac{3}{4\pi}\frac{M_{\rm halo}}{18\pi^2\rho_m}\right)^{1/3},
\end{equation}
with $18\pi^2$ is an overdensity at collapse redshift.
So the cross-section
mentioned before should not exceed $\sigma_{\rm max}=\pi R_{\rm
  vir}^2$. And the disc thickness when the latter is formed. From
\cite{OhHaiman_ApJ_2002} we have:
\begin{equation}
  R_{\rm Bondi}<H_{\rm disc}=\frac{c_s}{\sqrt{4\pi G \mu m_p n_0}}.
\end{equation}
The second limitation is the Eddington accretion rate (when
the falling matter rate is limited by the radiation push):
\begin{equation}
  \dot{M}_{\rm Edd}=\frac{1}{\epsilon}\frac{M_{\rm BH}}{t_{\rm
      Salp}},~~t_{\rm Salp}=\frac{c\sigma_T}{4\pi G m_p}\sim 450~{\rm Myr}.
\end{equation}
Typical value of radiation efficiency is $\epsilon\approx 0.1$. This
rate is also the upper limit of $\dot M_{\rm BH}$.

\begin{figure}
  \centering
  \includegraphics[angle=270,width=\linewidth]{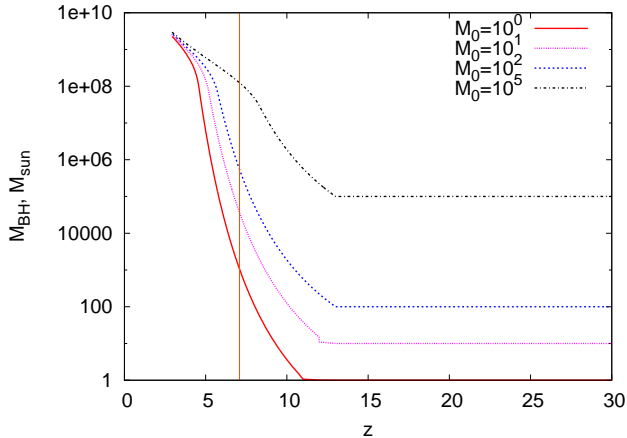}
  \caption{Dependence of a black hole mass vs redshift. The vertical
    line is at $z=7.085$.}
  \label{fig:z_Mbh}
\end{figure}

\begin{figure}
  \centering
  \includegraphics[angle=270,width=\linewidth]{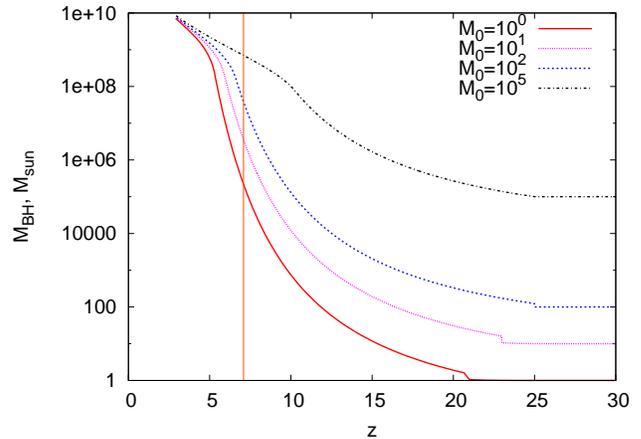}
  \caption{Dependence of a black hole mass vs redshift, 5-$\sigma$}
  \label{fig:z_Mbh5}
\end{figure}

The results of BH growth from 3-$\sigma$
fluctuations in the Press--Schechter formalism are shown on Fig. \ref{fig:z_Mbh}
for several initial BH masses: $M_0=10^0\div 10^2 M_\odot$ -- BH seeds
from massive PopIII stars, $M_0=10^5 M_\odot$ -- a direct collapse in
metal-free galaxies, see \cite{BegelmanEtAl2007}. 
The BH mass starts to grow significantly only after a baryon
disc creation. It quickly sets at the Eddington rate and further
decrease ($z\approx 5-10$)
of growth rate is connected with the $R_{\rm Bondi}<H_{\rm disc}$ criterium.
Our calculation shows that it is not so simple even though impossible
to obtain $M_{\rm BH}\approx 10^9 M_\odot$ at $z=7$. We could consider
as example the case of 5-$\sigma$ fluctuation in PS, see
Fig. \ref{fig:z_Mbh5}, and see that is hardly saves the situation.
The reason could be simply described in the following way: for
$3-\sigma$ fluctuations the active growth starts at $z\approx 11-13$
and is limited to the Eddington rate. From $z=13$ till $z=7$ we have
$\approx 340$ Myr, so $\exp(340/45)\sim 2\times 10^3$. So the BH should be
initially extremely massive to reach $10^9 M_\odot$.
These results are made with rough approximation, but we believe that they
reproduce the general properties of considered processes.

The mass of the quasar from \cite{MortlockWarrenEtAl_Nature_2011} is
explained in more sophisticated simulations by \cite{DiMatteo_ApJ_2012}
with account for hydrodynamics, star formation and feedback, for
details see \cite{SpringelEtAl_MNRAS_2005}. In those simulations
galaxies were created as initial conditions with nonzero angular
momentum. As was said earlier it requires careful analysis.

As a result of this section we could state that
creation of a supermassive black hole at high redshifts is a difficult
task. Though all known objects now are likely to be described by simulations, the
assumption of existence of a primordial black hole with large mass could simplify the
explanation and for more distant BHs, probably be discovered in future, it could be the only possible
explanation. The PBBH changes nucleosynthesis around itself. More
metals are created and it accelerates the process of local evolution,
and intensive growth could start earlier, at $z>13$ (for 3-$\sigma$,
from results of our calculations), this gives
more time for the BH growth with the Eddington rate.
This is the general reason why PBBHs are good candidates for dinosaurs.

\section{Conclusions}

The aim of the paper is to introduce the conception of ``cosmological
dinosaurs'' -- objects appeared long before the period of
intensive structure formation according to the standard model. The necessity for them could arise in
future: the observations are discovering now more and more distant galaxies and
quasars with no end in sight. At some moment this ``flow'' of newly
opened objects could contradict the accepted theory. We consider in
this case that the number of additional discoveries will be very
small, so only light extension of the theory is necessary.
In our work the appearance of such objects is connected with
primordial big black holes (PBBHs, we propose to introduce this term as
rich cosmological physics is connected with them). From our point of view these are the most natural and
simple extension: they start to form the structure around
themselves much earlier. The exact quantitative contribution could be
calculated in more sophisticated models, than presented in this paper
(e.g. with account for hydrodynamics) and is planned in future works. In the framework of our model
we have shown that modern high-redshift objects are on the edge of
the standard theory predictions. So the discoveries of dinosaurs or
effects of their previous appearance could be made in the nearest
future, introducing good probes for new physics in the early Universe.

SIG is partly supported by the project ``Development of
ultrahigh sensitive receiving systems of THz wavelength range for
radio astronomy and space missions'' in NSTU n.a. R.E. Alekseev, LCN.

\bibliographystyle{apsrev}
\bibliography{dinosaurs}

\end{document}